\documentclass{elsart41}
\usepackage{graphics}
\usepackage{graphicx}
\usepackage{amssymb}
\begin{document}

\begin{frontmatter}



\title{Thermoelectricity of EuCu$_{2}$(Ge$_{1-x}$Si$_x$)$_2$ intermetallics}
%

\author[CRO]{Veljko Zlatic\corauthref{Zlatic}},
\ead{name1@uni1.ac.xy}
\author[CH]{R. Monnier},\,
\author[US]{J. Freericks}

\address[CRO]{Institute of Physics, Bijeni\v{c}ka cesta 46, 
P. O. Box 304, 10001 Zagreb, Croatia}
\address[CH]{ETH H\"onggerberg, Laboratorium f\"ur 
Festk\"orperphysik,  Z\"urich, Switzerland} 
\address[US]{Georgetown University, Washingtonn D.C., USA}

\corauth[Zlatic]{Corresponding author. }

\begin{abstract}
Analyzing the heat and charge transport in a thermoelectric by 
transport equations, and assuming a simple  scaling between the 
current and the heat current densities, we evaluate ${Q}/{ S}$ and 
$ { Q}/\gamma T$ and find at low temperatures the 'universal  law' $q=N_A /N$. 
The deviations from the simple scaling law are also discussed and 
the results are used to analyze the effects of  chemical pressure on $q(x)$ 
in the EuCu$_2$(Ge$_{1-x}$Si$_x$)$_2$ and YbIn$_{1-x}$Ag${_x}$Cu$_4$  
intermetallic compounds. The characteristic temperatures of these 
compounds is greatly affected by doping, such that the ground state 
at $x=0$ and $x=1$ is completely different. 
A qualitative description of $q(x)$ can be obtained from the scaling law 
but the quantitative features require microscopic  modeling. We use for  
EuCu$_2$(Ge$_{1-x}$Si$_x$)$_2$ and YbIn$_{1-x}$Ag${_x}$Cu$_4$
the Anderson and the Falicov-Kimball model, respectively. 
\end{abstract}

\end{frontmatter}

This paper explains the thermopower $S(T)$ of EuCu$_2$(Ge$_{1-x}$Si$_x$)$_2$ 
intermetalic compounds. We discuss, first, the systematics which emerges from  
the experimental results as Ge concentration is varied from zero to 
one\cite{fukuda.03,hossain.04} and show that the properties of the paramagnetic 
phase follow from Kondo scatering on Eu ions, which fluctuate between 
two Hund's rule configurations. Assuming that Si-Ge substitution  gives rise 
to chemical pressure effect, which modifies the width and the position of an 
effective f-level representing the Eu ions, we show that  the NCA solution 
of the Anderson model gives the same features as seen in the data. 
(For details on the modeling of the rare earth intermetallics, and on 
our method of solution, see Ref. \cite{zlatic.05}.)

The thermopower of EuCu$_2$(Ge$_{1-x}$Si$_x$)$_2$ is 
characterized for $x \geq 0.8$ by a large peak centered at 
temperature $T_{max}\geq $ 150 K. The value of $S(T)$ at $T_{max}$ is 
large,  $S_{max} >  50 \mu V/K$, and above $T_{max}$ the thermopower 
falls-off slowly. At low temperatures $S(T)$ assumes a linear form and 
the slope $S(T)/T$ is small. The specific heat is featureless and $\gamma=C_V/T$ 
also is small. 
For $0.80 \geq x \geq 0.65$ the width of the thermopower peak is reduced 
and $T_{max}$ shifts rapidly to lower values as $x$ is reduced. 
Here, $S(T)$ changes sign for $T > T_{max}$.  
Since $S_{max}$ remains large, $S(T)/T$ is enhanced with 
respect to the $x \geq 0.8$ data.  In this concentration range $\gamma$ 
changes rapidly with $x$ and for $x<0.7$ assumes very  large values. 
The electrical resistance  $\rho(T)$ has a broad maximum which is shifted 
to lower temperatures as $x$ is reduced.  
The resistivity maximum occurs at $T\gg T_{max}$  but $\rho(T)$ 
drops off slowly and is quite large at $T_{max}$.  
The ground state is a Fermi liquid (FL) up to $x_c\simeq 0.65$ and  then  
becomes magnetic.  The specific heat and the electrical 
resistance show at $T_N$ a clear-cut anomaly but the AFM transition 
can be seen in the $S(T)$ data only  for $x < 0.6$. 
For $0.60 \leq x \leq 0.5$  the thermopower has a sharp cusp at $T_N$, 
i.e  the overall shape of $S(T)$ in these samples appears to be quite 
different from the shape seen in non-magnetic samples. 
However, if we assume $T_0 < T_N$ in this concentration range, 
we see that the magnetic samples  exhibit above $T_N$  the same features 
as the non-magnetic samples well above $T_0$. 
The size of the specific heat anomaly and the value of $T_N$ increase rapidly 
with $x$ up to about $x=0.5$\cite{fukuda.03}. 
Similarly, 
the electrical resistance also changes shape across the magnetic boundary.  
(See Fig. 6 in Ref. \cite{hossain.04})
That is, for $0.50\leq x \leq 1$, the transport properties of the paramagnetic phase  
follow the same pattern and seem to be due to the same scattering mechanism, 
regardless of the nature of the ground state which develops at low temperatures.  
For $x \leq 0.25$ the size of $S(T)$ becomes very small and is not affected 
by the AFM transition, which leads to a sharp anomaly in $C_V$ and also 
shows up clearly in the electrical resistance.  
(See Figs. 4 and 7 in Ref. \cite{hossain.04})

The initial thermopower slope of the paramagnetic samples changes by more than 
one order of magnitude. 

the Eu ions fluctuate between  the Eu$^{2+}$  and 
Eu$^{3+}$ configurations, as indicated by the photoemission data.

The Si doping reduces the unit-cell volume and leads to the mixing of the 
Eu$^{2+}$ and Eu$^{3+}$ configurations, as indicated clearly by the XPS 
data\cite{fukuda.03,hossain.04}. 
The AFM transition persists up to $x\simeq 0.65$ but already for 
$0.40 \leq x \leq 0.6$ the interconfigurational mixing leads to the 
Kondo effect with Kondo scale comparable to $T_N$. 
The behavior of these systems for $T>T_N$ can be explained 
by the NCA solution of the spin-degenerate Anderson model
with $S=7/2$ and using parameters such that $T_K < T_N$. Starting at high temperatures,  we find for $T\gg T_N$ 
a small negative thermopower and a large total entropy, 
such that $|\tilde q|\ll 1$\cite{zlatic.05b}. 
At temperature $T_0$ the thermopower changes sign and then 
continues to increase. $T_0$ is a monotonic function of $T_K$ 
but the relationship is not universal; for $T_K$ between 5 K and 20 K 
our calculations give $T_0\simeq 20 \ T_K$. (Note, the experimental result 
for $T_0$  is very sensitive to any additional contribution to ${ Q}$.)
The reduction of temperature gives rise to a partial screening of 
the Eu$^{2+}$ moment but  the magnetic entropy is not completely 
removed by Kondo effect, because the spin density wave (SDW) 
transition sets in at $T_N$. 
The neutron scattering data which might show the SDW are not available 
for EuCu$_2$(Ge$_{1-x}$Si$_x$)$_2$  but some evidence of a  SDW 
transition in a 'reduced-moment'  antiferromagnet is provided 
by recent neutron scattering data on CePd$_2$Si$_2$ compound at 
high pressure\cite{raymond.05}. 
(Unfortunately, the high-pressure thermopower and the specific heat data 
on  this compound are not available.)  
The SDW transition removes the magnetic entropy but the system remains 
in the metallic state and the specific heat below $T_N$ is temperature dependent. 
The thermopower tracks the entropy of the fermionic degrees of freedom 
and decreases linearly below $T_N$, so that ${ Q}(T)$ exhibits 
a clear cusp at $T_N$\cite{hossain.04}. 
The $T_N$ defined by this cusp, or by the discontinuity of $C_V$, 
increases with $x$ and reaches maximum at $x=0.55$. 
The size of the $C_V$-discontinuity is roughly independent of Si-concentration  
for $x\leq 0.55$ but ${ Q}(T_N)$ increases rapidly with $x$. 
For $x\geq 0.55$ the $T_N$ starts to decrease, the discontinuity of $C_V$ 
becomes weaker, and the cusp in ${ Q}(T)$ is gradually transformed 
into a rounded maximum. 
Thus, while $T_N$ shows a non-monotonic behavior,  $T_0$ and $T_K$ 
increase continuously with $x$, and $T_N/T_K$ changes from  
$T_N/T_K\gg 1$ for $x\simeq 0$ 
to $T_N/T_K\ll 1$ for $x\simeq 0.6$. 
For $T_N/T_K < 1$ and temperatures around $T_N$, the entropy 
has a large SDW contribution, such that $\tilde q <1$ but for $T\to 0$  
the universal behavior, $\tilde q\simeq 1$, should be recovered.
(Does one get $\lim_{T\to 0} q=1$ in CeAl$_2$, CeB$_6$, and other Kondo 
systems with $T_K > T_N$?)


%
The low-temperature specific heat and the thermopower change from 
${ Q}/T\simeq 12$ and  $\gamma\simeq 450$ for $x\geq 0.68$ to 
${ Q}/T\simeq 1$ and $\gamma\simeq 30$ for $x=1$, giving 
$q \simeq 1.5$ and $q\simeq 3$, respectively\cite{fukuda.03,hossain.04}. 
Thus, an increase in $T_K$ by more than one order of magnitude, 
and the complete change in the character of the ground state, 
has little effect on the low-temperature value of $q$.   
The slight increase of $q$ with the Si doping is most likely due to a reduction of 
the f-charge, which accompanies the enhanced mixed-valence character of 
the Si-rich samples and can be accounted for by Eq.\ref{resonant_SIAM}. 
The sharp drop of $\tilde q$ above $T_K$  indicates the crossover to 
weak coupling limit and the localization of the f-electrons, i.e., the change in $V_F$. 
\vspace*{10mm}
\begin{figure}[!ht]
\begin{center}
\includegraphics[width=0.45\textwidth]{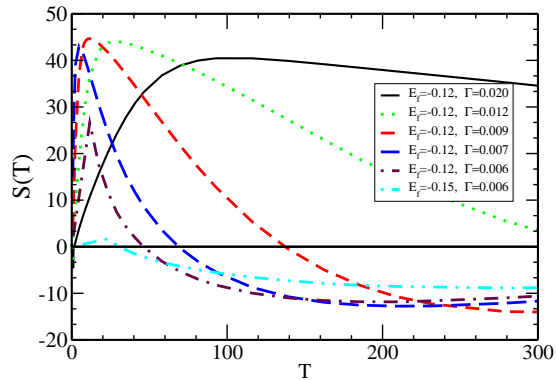}
\end{center}
\caption{The NCA result for the thermopower plotted versus temperature 
for $E_f$=-0.12 and for various hybridization widths $\Gamma$, 
as indicated in the figure. }
\label{fig1}
\end{figure}
\section*{Acknowledgement}
We acknowledge the financial support from the National Science Foundation 
under grant number DMR-0210717. One of us (V. Z.)  acknowledges the 
Mercator-Guestprofessorship award from the German Research Foundation.
We are grateful to B. Coqblin for many stimulating discussions.


\begin{thebibliography}{99}

\bibitem{fukuda.03}
S. Fukuda et al.,  Journal Phys. Soc. Japan {\bf 72} (2003) 3189. 

\bibitem{hossain.04}
Z. Hossain et al.,  
Phys. Rev. B {\bf 69} (2004) 014422. 

\bibitem{zlatic.05}
V. Zlati\'c and R. Monnier, 
Physical Review B{\bf 71} (2005) 165109.



\end{thebibliography}
\end{document}